# Propelling Innovation to Defeat Data-Leakage Hardware Trojans: From Theory to Practice

Kevin Kwiat
CAESAR Group
Sarasota, Florida, USA
kevink@caesargroup.org

Jason Kulick
Indiana Integrated Circuits
South Bend, Indiana, USA
jason.kulick@indianaic.com

Paul Ratazzi
AFRL, Information Directorate
Rome, New York, USA
edward.ratazzi@afrl.af.mil

*Abstract*—Many design companies have gone fabless and rely on external fabrication facilities to produce chips due to increasing cost of semiconductor manufacturing. However, not all of these facilities can be considered trustworthy; some may inject hardware Trojans and jeopardize the security of the system. One common objective of hardware Trojans is to establish a side channel for data leakage. While extensive literature exists on various defensive measures, almost all of them focus on preventing the establishment of side channels, and can be compromised if attackers gain access to the physical chip and can perform reverse engineering between multiple fabrication runs. In this paper, we advance (from theory to practice) RECORD: Randomized Encoding of COmbinational Logic for Resistance to Data Leakage. RECORD is a novel scheme of temporarily randomized encoding for combinational logic that, with the aid of Quilt Packaging, prevents attackers from interpreting the data.

## I  Introduction

A continuing trend of US semiconductor companies is to outsource their device fabrication (i.e., fabless manufacturing) to specialized, yet untrusted, foundries that are overseas. An international electronics manufacturing company with suspected ties to a foreign government has brought the issue of malicious hardware to the forefront [1]. Meanwhile, the insider threat undermines the security of on-shore production of microelectronics. The "buyer beware" mantra has taken on a more sinister meaning for procurers of microcircuits. In the realm of cybersecurity, many regard confidentiality as paramount because once a secret is lost it is irrecoverable. Spurred by these conditions, the U.S. Air Force patented inventions [2]-[4], subsequently licensed by Indiana Integrated Circuits (IIC), that undermine in-situ data-leakage hardware Trojans. These inventions were advanced through participation in the AFWERX's Microelectronics Design Challenge. They represent a last-line of defense in a dogged determination to defeat such an attack by fighting-through it.

## II  Framework for Fighting-Through

Fighting-through a hardware Trojan leaking confidential data off a chip is performed by RECORD (Randomized Encoding of COmbinational Logic for Resistance to Data Leakage). RECORD's framework has two main components: 1) it introduces randomness in the information processing; and 2) it employs split manufacturing. Both of these components will be delved into; however, it is worthwhile at this point to mention that joining the result of having split manufacturing is a very low-loss, high-bandwidth chip-to-chip integration technology called Quilt Packaging (QP) [5].

### A. Randomization

RECORD introduces randomization by combining a circuit's primary inputs, $x$'s, with a random number generator's output bit, $r$, to produce a temporary, random value $t$:

$$t = x \oplus r \tag{1}$$

With all input signals sharing the same random bit, any given combinational function, $f$, can be converted as follows:

$$f(x_1, x_2, \ldots) \rightarrow (f(t_1 \oplus r, t_2 \oplus r, \ldots) \oplus r, r) \tag{2}$$

Applying Shannon's Expansion Theorem to (2) yields:

$$f(t_1 \oplus r, t_2 \oplus r, t_3 \oplus r, \ldots) \oplus r \tag{3}$$
$$= \overline{r} f(t_1, t_2, t_3 \ldots) + r \overline{f(\overline{t_1}, \overline{t_2}, \overline{t_3}, \ldots)}$$

Note that the last expression in (3) is that of a two-input multiplexer with inputs $f$ and $\overline{f}$ and select $r$. Fig. 1 shows the translation of this expression into a hardware representation.

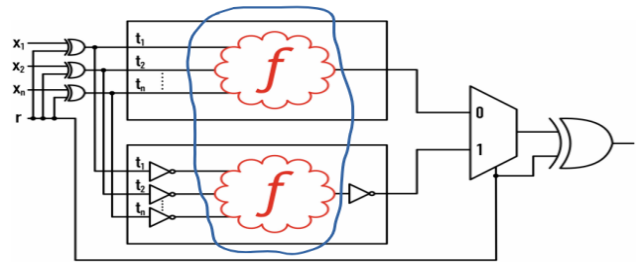

Fig. 1. RECORD's digital logic (outsourced replicated functionality *f*)

On the left of Fig. 1, RECORD uses random bit *r* to figuratively "flip a coin" - thus creating randomly inverted input streams for the replicated *f*s. Although some of the resultant outputs of these *f*s must be wrong, they are far from worthless because they usher-in a "shell game" effect from which the multiplexer on the right chooses, based on *r*, the correct bit in preparation for the final, desired output. The XOR connected to the multiplexer's output performs this final step: its other input is *r* so the XORing nullifies whatever effects *r* had on the datapath through the circuit. If the *f*s are outsourced such that any data-leakage hardware Trojans are restricted to the area represented by the area enclosed in blue in Fig. 1, then the data leaked has been randomized by *r*. This is true as long as the Trojans cannot obtain or derive random bit *r*.

*B. Split Manufacturing*

Restricting a hardware Trojan's access to *r* is done by introducing a split manufacturing step in RECORD. Whereas the *f*s are outsourced, the remaining logic is produced in a trusted facility – such as a DoD Trusted Foundry [6]. This logic is composed of the following: a random number generator, XOR gates, multiplexers, and inverters. Together, they form a set of what is referred to as trusted I/O modules. This set is universal to RECORD and therefore the I/O modules are generic for all RECORD implementations. After the trusted facility obtains the outsourced *f*s, QP is applied to join them to the trusted I/O modules to produce a completed RECORD design as conceptually represented in Fig. 2.

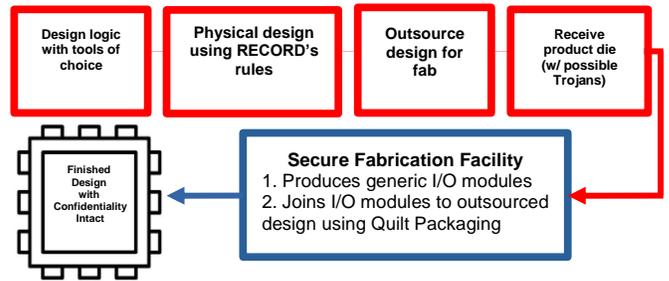

Fig. 3. RECORD's design flow

The resultant chip area of a RECORDized design, with requisite replication of *f,* is the primary cost. An 8-bit Advanced Encryption Standard (AES) Substitution Box (Sbox) was synthesized and laid-out as a measure of RECORD's expected design costs [7]. Table I shows the maximum design penalties (in terms of power, area, and delay) for this sample RECORDized design.

TABLE I.

| Attribute Increased | Example of RECORD's Costs |
|---|---|
| Power | 3.4x for Dynamic, and 2.19x for Leakage |
| Area | 2.4x |
| Delay | ≤ 11% |

It is possible to reduce the costs in Table I (albeit slightly) by reducing the number of device inputs that are randomized. Reductions that were achieved for the sample design of Table I were reported in [7]. Reducing the number of inputs that are randomized may be warranted if *f*'s application is not negatively impacted by doing so. For example, a coded input to *f*, such as ASCII, may have fewer than 9 bits randomized and still retain a desired degree of confidentiality.

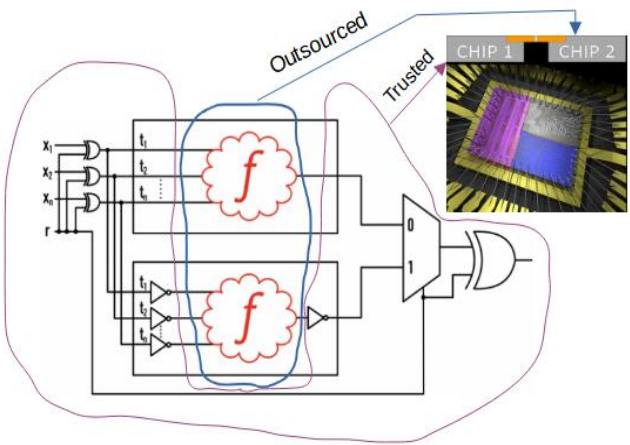

Fig. 2. QP'd *f*s (blue area) with I/O modules (lavender area)

### III. DESIGN CONSIDERATIONS

RECORD's 2-step process of introducing randomness in the information processing and split manufacturing restores confidentiality (i.e., thwarts data-leakage) when trust was surrendered by outsourcing the *f*s.

Having described what RECORD is, it is worthwhile to say something about what RECORD is not. RECORD is not a completely new way to design digital logic; instead, designers are free to use the CAD/CAE tools of their choice for logic design. It is assumed that duplication of the *f*s, should, during logic design, be straightforward. RECORD's rules for physical design, however, differentiate from a typical ASIC design flow because the outsourced die must have QP interfacing. This is illustrated in Fig. 3. Incidentally, QP is not limited to connecting two dice; instead, the RECORDized design in Fig. 2 is an example by showing a third die (i.e., the gray rectangle) being QP'd to the *f*s (in blue) and the I/O modules (in lavender).

Although the RECORD acronym specifically calls out combinational logic, RECORD is not limited to combinational logic; instead, a sequential logic version of RECORD exists [4], [8]. In such designs, the flip-flops' values must be synchronized with changing values of *r*; therefore, like the generic I/O modules which contain *r*, the memory elements must come from a trustworthy source and be QP'd to the outsourced core of combinational logic composing the *f*s. While other details of sequential RECORD can be found in [4], [8], Fig. 3 represents a block of prefabricated registers (i.e., the gray rectangle) having been quilted to the outsourced core. However, since quilting is only done on the perimeter of the outsourced core, the I/O between it and the register block can quickly become overwhelmed with an increase in the number of registers. In order to keep the random bit, *r*, secure, the split manufacturing process is modified: instead of using QP to combine the trusted and untrusted portions of the chip, Sequential RECORD can use a secure top tier and 3D manufacturing to combine the two tiers using Through Silicon Vias (TSVs) [9].

## IV. Physical Instantiation

Quilt Packaging and 3D manufacturing with TSVs are among the physical manipulations possible when designing with RECORD. There is another manipulation that came to the forefront with RECORD. The DoD Executive Agent (EA) for Printed Circuit Board (PCB) and Interconnect Technology, supported a hardware demo of RECORD at the PCB level where the $f$s perform enhancement of UAV-obtained imagery. This demo called not for quilting chips, but for quilting boards. A demonstration of fighting-through a data-leakage hardware Trojan at the PCB level was warranted because, unlike chips, once boards leave their factory they remain susceptible to rogue devices being planted on them. Fortunately, IIC et al, created a method that elevated QP from chips to PCBs [10].

Fig. 4 shows the demo set-up for RECORD at the PCB level. In the lower right are the quilted boards where the left half has the trusted generic I/O modules and the right half represents the untrusted, outsourced portion of RECORD. The FPGA on the right half contains the image enhancement functionality – along with always-on, full-access data-leakage hardware Trojans.

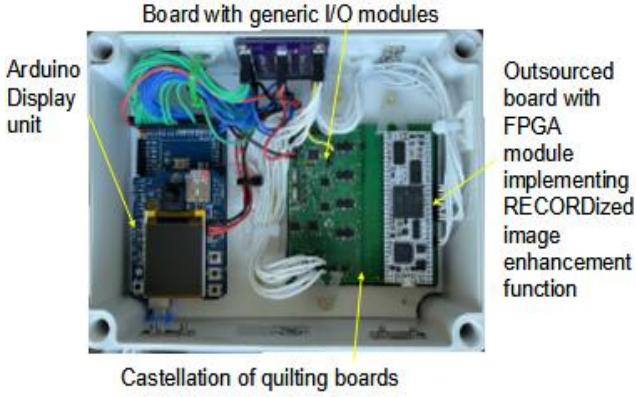

Fig. 4. PCB RECORD Demonstration

Prior to mapping the dual, combinational logic of the image enhancement functionality to the FPGA, simulation of the RECORD's design revealed a shortfall: the displayed output of the demo showed that the data leaked by the Trojans to be insufficiently randomized. Although RECORD induced obfuscation in the processed pixels, the resultant image's objects could be faintly discerned.

To overcome this shortfall, we turned to the mathematical basis of RECORD where a second random bit, $r2$, is introduced using (3) as a baseline. The resultant equation is:

$$f(t_1 \oplus r_1, t_2 \oplus r_2, t_3 \oplus r_1, \ldots) \oplus r_1 \\ = \overline{r_1 r_2} f(t_1, t_2, t_3, \ldots) + \overline{r_1} r_2 f(t_1, \overline{t_2}, t_3, \ldots) \\ + r_1 \overline{r_2} f(\overline{t_1}, t_2, \overline{t_3}, \ldots) + r_1 r_2 f(\overline{t_1}, \overline{t_2}, \overline{t_3}, \ldots) \quad (4)$$

Note that the righthand side of (4) is that of a four-input multiplexer with select bits $r_1 r_2$ and with four instances of $f$ (as shown in (4)) serving as the source for the multiplexer's inputs. Although this modification dictated that we quadruple the number of instances of $f$, the demonstration's FPGA had ample logic resources to accommodate this added overhead.

## V. Demonstration Results

Following the loading of the FPGA with the four instances of $f$, the demo was ready for operation. The demo's display unit rendered the following: the initial (unprocessed) UAV-obtained image; the image enhanced through noise removal; and the image leaked by the hardware Trojans.

Fig. 5 shows the initial UAV-obtained image with noise present. The image's lack of clarity due to noise is apparent.

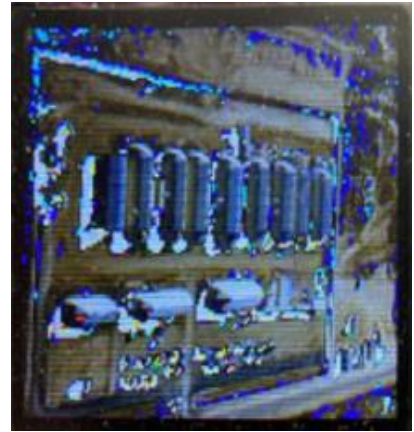

Fig. 5. Original Image

Fig. 6 shows the demo's output of RECORD for the bona fide user.

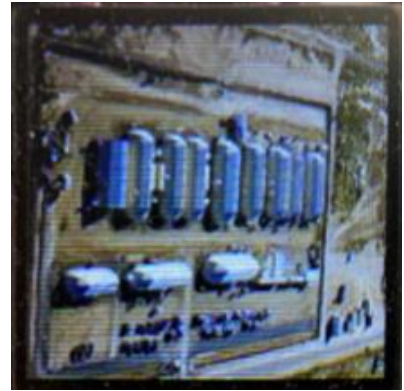

Fig. 6. Enhanced Image

For the demo's in-situ hardware Trojans, we gave them full-access to all four instances of the $f$s in the FPGA; thus mimicking the outcome an attacker with access to the design, implementation, and physical instantiation of the $f$s. Fig. 7 shows what those Trojans are able to leak.

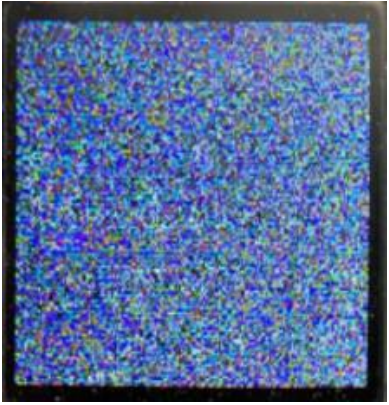

Fig. 7. Trojan-leaked Image

## VI. INSIGHTS DERIVED

The demo provides compelling visual evidence that RECORD defeats confidentiality attacks from the presence of malicious hardware. The payoff is that a smart buyer can harness the global design and production capabilities while being assured that the trust in confidentiality surrendered to those untrusted sources of supply can be restored. As if affirming that "insights incite change", the demo's results encouraged us to investigate extensions of RECORD.

### A. Fault-Tolerant RECORD

The presence of RECORD's dual $f$s, as shown in Fig. 1, lend themselves to augmenting so that they can simultaneously provide Trojan blocking together with tolerance of a single transiently faulty $f$ that occurs due to natural phenomena. Such faults would occur in single-event upset (SEU) environments. A tension between fault tolerance (for reliability) and security arises here: if introducing redundancy for fault tolerance purposes also multiplies a secret, then it is contrary to assuring confidentiality. When operating RECORD in a fault-tolerant manner, some of the $f$s are identically executing. The data being processed by those $f$s, being identical, would, after being leaked, reveal the data's importance to the attacker who then could more readily interpret the information. To avoid this possibility, the $f$s cannot share the same substrate. That is, unlike the RECORD designs previously described, in a fault-tolerant RECORD the $f$s must be individually quilted together or packaged as individual die in a hybrid integrated circuit. Doing so allows prevention of communication between the Trojan-infected $f$s. Being unable to communicate, the Trojans' actions become uncoordinated and the data they leak more disorganized. This reduces the risk associated with redundantly-executing $f$s in RECORD. Furthermore, it is assumed that the Trojans in the independent $f$s are solely seeking to leak data, so they would not intentionally disrupt operation by inducing faults.

Fig. 8 shows a design for fault-tolerant RECORD where a redundant, third $f$ appears at the bottom of Fig. 8.

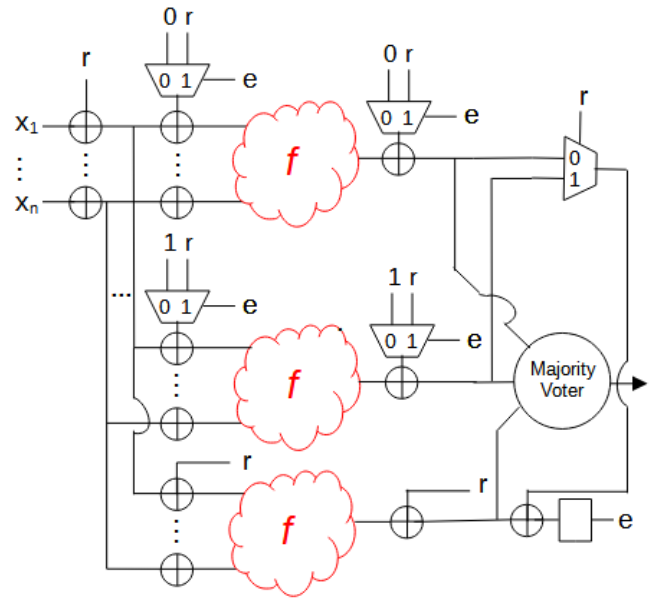

Fig. 8. A Fault-Tolerant RECORD

The third $f$ has two objectives: fault detection; and, when a fault is detected, fault masking. Therefore, the circuit of Fig. 8 operates in two phases.

- In the first phase, the top two $f$s operate in normal RECORD fashion while the third $f$ operates redundantly by mirroring the behavior of the $f$ selected by $r$. In this way, the third $f$ operates as a pair-and-spare. The XOR at the third $f$'s output compares the spare's output with the output selected from the other $f$s., When a miscompare occurs, the XOR sets the error signal, $e$, of a flip-flop to "1". To prevent an error from propagating to the output, it is assumed that the circuit's output is buffered (buffer not shown) and the values of the circuit's $x$ inputs and $r$ are saved in order to correct the current fault's effects.

- In the second phase, where $e$ has been set to "1", all three $f$s operate based on r (being replayed together with the $x$ inputs) and hence form triple-modular redundancy (TMR). From the TMR, the voter of Fig. 8 produces, by fault masking, a fault-free output that replaces the previously buffered faulty value. The signal $e$ is then set to "0" and first-phase operation returns.

Again, the above two phases assume a transient fault in an $f$. If, instead, a permanent fault was to occur, then the circuit of Fig. 8 must be modified to allow purging of the permanently faulty $f$; otherwise, the circuit remains in TMR mode where the $f$s identically process those inputs. The sameness of the resultant data could likewise reveal its importance to the attacker. Purging as a safeguard in fault-tolerant RECORD underscores the challenge of combining reliability and security both reliably and securely.

### B. System-on-Chip Considerations

In a system-on-chip (SoC) it would likely be prohibitive to employ RECORD over the entire design; yet QP allows for selectively applying RECORD. Given RECORD's support for confidentiality, an SoC's crypto-core would be a suitable consideration. If this core is outsourced and QP'd to generic I/O modules, the resultant chiplet can then be further QP'd to an un-RECORDized SoC's components. In this way, a SoC designer can garner the benefits of RECORD without having to absorb the technique's overhead across the entire design.

*C. Leave-Behind Applications*

RECORD's division of a design into trusted and untrusted regions has opened a straightforward approach to address impacts to the supply chain posed by "leave-behind" applications. The UAV-obtained imagery of Section V illustrates the possibility of a downed drone becoming the source of adversarial reverse engineering of its microcircuits. Assuming that the adversary can then gain complete knowledge of the chip, future releases of that design are in jeopardy because whatever design-for-security (DFS) methods had been applied are now compromised. The impacts to the supply chain can be dramatic: preventing insertion of future hardware Trojan calls for revamping the chip's DFS - effectively burdening the supply chain with a new design. However, when reverse engineering of a RECORD design occurs, the adverse effects to the supply chain can be substantially lessened. Quilting a new generic I/O module (that changes the generation of $r$) to the outsourced design does not diminish RECORD - nor does it require altering the outsourced design, but allows it to remain unchanged for future hardware releases.

*D. Admitting Other Hardware Security Techniques*

RECORD's dependence on QP has physical design implications. Although the components being QP'd must have the proper interface, our investigations revealed that QP is not a limiting factor in applying the following DFS techniques to the $f$s: split manufacturing (i.e., further dividing the production of each individual $f$), obfuscation, and layout camouflaging. Affording designers the freedom to employ these other DFS techniques allows for envisioning a chip possessing defense-in-depth with RECORD serving as the last bastion for confidentiality against in-situ, data-leakage hardware Trojans.

*E. Deterring Other Hardware Trojan Types*

Hardware Trojans that attack a chip's integrity or availability are also affected by RECORD. When the attacker aims to trigger such Trojans once $f$ encounters a predetermined sequence of inputs, RECORD's randomization acts to disrupt the sequence. Even though the attacker knows RECORD is being deployed, randomization may eventually thwart triggering of the Trojan. Furthermore, knowing of randomization's presence complicates the attacker's initial strategy for hardware Trojan creation.

## VII. CONCLUSION

The irreversibility of having lost confidential data is exacerbated by the stealthiness of data-leakage Trojans whose presence in hardware can go undetected – indefinitely. Their undetectability propelled finding a way to fight-through them in both microcircuits and PCBs. RECORD's area and power costs may be acceptable for new designs having sufficient headroom. For upgrading legacy systems, the past influence of Moore's Law (where transistor density roughly doubled every previous 18 months) can offset RECORD's costs - being an incentive to RECORDize those older systems' chips and PCBs. Moreover, such an endeavor synergizes with the US government's efforts to act upon how the current overreliance on China for less-advanced chips is a threat to national security [11].